\documentstyle[12pt,aasms4]{article}

\begin{document}

\title{Are optically--selected QSO catalogs biased ?}
\author{Ignacio Ferreras\altaffilmark{1},
Narciso Ben\'\i tez\altaffilmark{1} \and
Enrique Mart\'\i nez-Gonz\'alez}
\affil{IFCA, CSIC--Universidad de Cantabria, Facultad de Ciencias, Avda.
los Castros s/n, 39005 Santander, Spain}

\altaffiltext{1}{Dpto. de F\'\i sica Moderna. Universidad de Cantabria,
Fac. Ciencias. Av. los Castros s/n. 39005 Santander, Spain}
\authoremail{ferreras@ifca.unican.es}

\begin{abstract}
A thorough study of QSO--galaxy correlations has been done
on a region close to the North Galactic Pole using 
a complete subsample of the optically selected 
CFHT/MMT QSO survey and the galaxy catalog of Odewahn and Aldering (1995). 
Although a positive correlation between bright QSOs and galaxies 
is expected because of the magnification bias effect, none is detected. 
On the contrary, there is a significant ($>99.6$\%) anticorrelation 
between $z<1.6$ QSOs and red galaxies on rather large angular distances. 
This anticorrelation is much less pronounced for high redshift $z>1.6$ QSOs, 
which seems to exclude dust as a cause of the QSO underdensity. 
This result suggests that the selection process employed 
in the CFHT/MMT QSO survey is losing up to $50\%$ of low redshift 
$z<1.6$ QSOs in regions of high galaxy density. The incompleteness 
in the whole 
$z<1.6$ QSO sample may reach 10\% and have important consequences 
in the estimation of QSO evolution and the QSO autocorrelation function. 

\end{abstract}

\keywords{cosmology: gravitational lensing --- large-scale structure of 
universe, quasars: general, surveys}

\section{INTRODUCTION}

There has been recently a flurry of papers searching for QSO--galaxy
associations caused by gravitational lensing. The masses of foreground
galaxies act as lenses on the light beams from background quasars. These
QSOs will be affected in two ways: on one hand the beam will be focused,
increasing the brightness of the source resulting in a higher population
of quasars close (in angular distance) to foreground galaxies. On the other
hand, the lensing will also magnify the solid angle, thereby decreasing
the number density of QSOs. The tug between these two effects will determine
the correlation between foreground and background sources. If the slope
of the QSO number counts function is steep enough, there will be many
faint quasars to allow an overdensity and thus a 
positive correlation, whereas a flat slope
implies there are not many faint QSOs to balance the increase of solid
angle, resulting in a negative correlation function (Narayan 1989; 
\markcite{na89} Schneider, Ehlers \& Falco 1992\markcite{sef92}). 
If the background sources are flux-limited in two uncorrelated wavebands 
( e.g. radioloud QSOs) the total effect will be determined by the sum 
of the number counts slopes in both bands. This is called 'double 
magnification bias' (Borgeest, Von Linde \& Refsdal 1991\markcite{bo91}).
So far, the observational results on large scales suggest a 
positive correlation for radio--selected QSO samples 
(Bartelmann \& Schneider 1994 \markcite{bs94}; Ben\'\i tez \& 
Mart\'\i nez-Gonz\'alez 1995 \markcite{bm95}) and null 
correlation or even anticorrelation for radio quiet quasars.
Although the magnification bias may produce anticorrelations for faint 
QSOs (as the ones in Boyle, Fong \& Shanks 1988\markcite{bo88}), they 
cannot explain the negative correlations found in Ben\'\i tez \& 
Mart\'\i nez-Gonz\'alez (1997) \markcite{bm97} or 
Ben\'\i tez, Ferreras \& Mart\'\i nez-Gonz\'alez (1997) \markcite{bfm97}, 
where LBQS QSOs with ($B_J\lesssim 18.7$) are considered. 
It is thus more plausible to interpret these anticorrelations 
as caused by dust extinction in foreground clusters 
(Romani \& Maoz 1992\markcite{rm92}) or selection biases due to 
the difficulty of identifying quasars in crowded fields 
(Maoz 1995\markcite{ma95}). 

It is interesting to confirm the existence of these biases and clarify 
their origin as they may have important effects in the estimation of the QSO 
autocorrelation function or the QSO evolution. It is also remarkable that 
the conclusions obtained from gravitational lens surveys employing radioquiet 
quasars may be seriously affected by these selection effects. 
Along these lines we decided to explore the QSO-galaxy correlation 
function for the grens-based CFHT/MMT survey. 
Hartwick \& Schade (1990)\markcite{hs90} conclude that this 
survey is probably the most efficient method of detecting faint quasars by 
slitless spectroscopy. Apart from being reputedly free of biases, 
this catalog is also interesting for our purposes as it covers a 
large range of QSO magnitudes and redshifts. 

The galaxy catalogue of Odewahn \& Aldering (1995)\markcite{od95}, 
is also very appropriate for our purposes as it offers a photometrically 
homogeneous sample on a large region of the sky and contains colour 
and rough morphological information for the galaxies. As it was shown 
in Ben\'\i tez \& Mart\'\i nez-Gonz\'alez (1995) \markcite{bm95}, 
selecting galaxies by their colors may produce dramatic effects in 
the behavior of the QSO-galaxy correlation function.

Section 2 shows the details of the data that made possible this work. 
Section 3 deals with QSO--galaxy cross-correlations, using two
different approaches. The next section extracts the information 
on small angular distances using again two independent estimators.
In the final section we discuss the implications and make an attempt 
at interpreting these results. It is concluded that bias effects in 
optically selected QSO catalogs may seriously affect the outcome of 
many studies based on this kind of quasars.

\section{DATA}

The statistical analysis done in this paper is based on two complete
sets of galaxies and quasars respectively. The galaxy
catalog compiled by Odewahn and Aldering (1995) \markcite{od95} 
was obtained from the APS scans of nine  POSS-I photographic plates 
around the North Galactic Pole, covering an area of 289 deg$^2$. 
In the calibration process, extra care was taken to avoid 
plate-to-plate variations in the photometry. The catalog lists $O$ 
and $E$ photographic magnitudes as well as a concentration index. This feature
makes the sample especially interesing since it is possible to give a 
rough estimate of the morphology of each galaxy taking its $O-E$ color and
concentration. A suitable choice of subsamples allows 
an estimation of the distribution of galaxies on large scales. 
The region shows a conspicuous degree-sized structure
including the Coma cluster as well as many filaments. The complete 
list comprises 36402 galaxies.

The quasar catalog comes from the CFHT/MMT survey \markcite{cr89} 
(Crampton, Cowley \& Hartwick 1989) based on grens spectra taken at the CFHT,
using III-aJ photographic plates.
Candidates were selected based on visual inspection searching for
prominent emission lines (mainly \ion{C}{3}, \ion{C}{4}, \ion{Mg}{2}, 
and Ly-$\alpha$), or objects with unusually blue continua. A high
percentage of confirmed candidates makes this survey one of the most
efficient ones.
There is supposed to be no evidence of significant
incompleteness in the range $0.3<z<3.4$ (Hartwick \& Schade 1990) 
\markcite{hs90}. Crampton et al. surveyed several fields one 
of which --- Field 1338+27 --- is complete in a $\sim$ 5.5 deg$^2$ 
region that overlaps with the galaxy sample and comprises 160 QSOs. 
Figure 1 shows these quasars superimposed on a smoothed image of the 
galaxy density. This region of the 
sky is quite remarkable because of a degree--sized filament of red 
galaxies stretching along the southern portion of the field. These two 
catalogs are thus ideal to explore effects of foreground structures on 
the number density and distribution of background QSOs. 

In the next section we search for possible correlations both at
small and large scales, using the complete sample of quasars and galaxies as
well as several subsets listed below:
\begin{itemize}
\item [$\bullet$] Red ($O-E\geq 1.5$) and Blue ($O-E<1.5$) Galaxies.
\item [$\bullet$] Concentrated ($c_{31}\geq 2.4$) and 
		   Non-Concentrated ($c_{31}< 2.4$) Galaxies.
\item [$\bullet$] Bright ($B_j\leq 19.6$) and Faint ($B_j>19.6$) Quasars.
\item [$\bullet$] High-z ($z>1.6$) and Low-z ($z\leq 1.6$) Quasars.
\end{itemize}
The average values were the ones chosen to split the population of 
QSOs and galaxies so that each subsample has roughly the same number
of objects. Just by eyeballing the subsamples we clearly see the 
large scale structure made of red and concentrated galaxies 
(i.e. old ellipticals). 
It is quite instructive to compare the results from Odewahn \& Aldering 
(1995): The mean $O-E$ color in the filamentary region is $1.53\pm 0.02$ 
compared to a bluer $1.32\pm 0.01$ in the interfilament area. Besides, the 
galaxies in the filaments are more concentrated, with a $c_{31}$ index of
$2.226\pm 0.006$ in contrast with $2.153\pm 0.005$ in the outside 
of the filaments.

\section{QSO--Galaxy cross-correlations}

This paper aims at estimating the cross-correlation between QSOs and galaxies.
Such correlations give important information about background quasars 
being lensed by foreground galaxies. On the other hand, the QSO sample 
is assumed to be complete, which allows us to infer properties on their 
large scale distribution. 
The features of this sample make the analysis totally different from that 
of previous work. The galaxies are not weighed equally: the ones closer 
to the field center will be added more times when computing
correlations than the ones in the outside. Hence, a comparison between
the actual QSO catalog and a random QSO sample is essential to eliminate
possible ``contaminations'' in the correlations coming from galaxy--galaxy
clustering. In this section we explore the cross-correlation using two 
different methods: the correlation function, and a robust treatment
comparing QSO and galaxy number densities.

\subsection{Correlation Function}
The correlation function gives the excess probability of finding a 
quasar--galaxy pair over that of a Poisson distribution. In this paper,
care has to be taken in order to define a suitable correlation estimator.
If the sample were big enough so that the density achieved homogeneity
at large angular distances, then we could use the ``standard'' definition
that normalizes the pair number with the area. Instead, we have to use
a different normalization so that the function goes to zero at large 
angular distances.
We count the number of galaxies inside rings centered at each quasar;
then we compute an area correction by throwing 30000 random points 
homogeneously distributed over the QSO survey region,
and then count the number of the random points which fall inside the
rings. If $n_g(\theta_i)$ represents the number of galaxies in a distance
range $\theta_i \le \theta < \theta_i + \Delta\theta$ added for all the
QSOs in the sample; and $n_r(\theta_i)$ is the number of random points 
inside the same ring, then we define a normalized pair ratio:
\begin{equation}
{\cal N}_g(\theta_i) \equiv {n_g(\theta_i)\over n_r(\theta_i)} 
\times {\sum_i n_r(\theta_i)\over\sum_i n_g(\theta_i)} 
\end{equation}
A random distribution of points would give ${\cal N}_g(\theta_i)= 1$, hence
its deviation from $1$ is the correlation function. However, this function
is extremely sensitive to the galaxy population, i.e. if there is a strong
gradient in the number of galaxies across the field, then the correlation 
function will yield a bogus association caused by this gradient.
The way out of this is computing the normalized pair ratio again for a 
simulated sample, which could be just a random distribution with the
same number of QSOs as the original one. Instead, we decided to 
choose a slightly different comparison catalog: We divided the region
in $6\times 6$ boxes ($\sim 25^\prime \times 22^\prime$), shuffling and
rotating them 100 times.
This method will yield error bars that are more
realistic than the ones obtained by throwing random points all over the
QSO region. If we write the pair ratio for the random sample as
${\cal N}_g^{\rm rnd}(\theta_i)$ then the correlation function is:
\begin{equation}
\omega_{\rm qg}(\theta_i) = 
{{\cal N}_g(\theta_i)\over {\cal N}_g^{\rm rnd}(\theta_i)} - 1
\end{equation}
The correlation function is shown in Figure 2 for the complete 
sample as well as for partial subsamples taking red and blue galaxies 
and high and low redshift quasars. We used 200 simulated samples in each
run to compute the correlation function and the error bars. The largest
difference between subsamples appears when taking high and low redshift
QSOs. The $z<1.6$ population has a strong anticorrelation with 
galaxies that is more pronounced when restricting the foreground sample
to red ($O-E>1.5$) galaxies, i.e. when dealing with the (elliptical) galaxies
that dominate the gradient in number density. This paucity of quasars can
be even checked by eye taking a look at Figure 1. To quantify this
result, we applied a Spearman rank correlation test comparing the value of
the correlation function against distance. We tried different cutoffs
($\theta_{\rm cutoff}=120^\prime ,150^\prime ,180^\prime ,200^\prime$)
and all of them agreed with a confidence level $\approx 100$\% for the
red-galaxy/low-z-QSO subsamples. The bump at around $80^\prime$ is caused
by the filament of galaxies in the low declination area. This bump gives a 
coarse estimate of the distance between the ``centers of mass'' of both
quasar and galaxy catalogs.

\subsection{QSO \& Galaxy density diagrams}
In order to double check this result, we used another method to estimate
the negative cross-correlation found. The most straightforward way of 
checking whether there is a correlation 
on large scales is analysing the density of quasars as a function
of galaxy density. We binned the region in $n\times n$ boxes 
(in R.A. and Declination) counting QSOs and galaxies inside
these boxes and sorting them with respect to the galaxy counts.
Then we checked the results rebinning them as well as using different 
values of $n$ to make sure the result was stable. 
A suitable value of $n$ is given by the average 
quasar--quasar separation, roughly
10 arcminutes. Figure 3 shows several $q = \rho_{\rm QSO} /\langle 
\rho_{\rm QSO}\rangle$ vs. $\Delta\rho_{\rm g} /\langle\rho_{\rm g}
\rangle$ plots for $n=15$ ($10^\prime\times 9^\prime$ boxes), where
the average galaxy density is computed in the region where the QSO 
survey is complete.
The total $15\times 15$ points were rebinned for a better visualization.
A Rank correlation applied to the complete sample of QSOs and galaxies
with no rebinning gave a statistic which is 1.7$\sigma$ away from the
expected value for a homogenous distribution, implying a confidence
level of 95.5\%. 

However, the result is much more significant if we separate the list in 
high and low redshift quasars. Notice the remarkable difference between 
these two populations in the bottom panels of Figure 3.
Taking the red-galaxies/high-$z$-QSOs subsample we find the rank 
correlation statistic
to lie 2.6$\sigma$ away from the prediction for a random sample, giving a
confidence level of 99.5 \%. Furthermore, we wanted to check for possible
deviations from a Gaussian profile and performed the same treatment using
1000 random QSO catalogs. Only 4 out of 1000 gave a higher statistic than
the real sample, thereby yielding a 99.6\% confidence level, in agreement
with the result obtained using the correlation function. Besides, the
high redshift population behaves quite in the opposite way, suggesting
only at a 92\% confidence level a positive correlation with galaxy density.
Anyway, the remarkable difference found between high and low redshift
quasars is a clear hint that strong biasing might be present in one
of the selection methods assumed to be most unbiased so far. This effect
could be explained by the fact that high-$z$ quasars are detected searching
for their prominent Lyman-$\alpha$ emission line whereas lower redshift 
candidates shift this line out of the spectral range of the photographic
emulsion and so their detection is based on weaker --- thereby harder 
to detect --- emission lines such as \ion{C}{4}, \ion{Mg}{2} and so on.

\section{Sub-degree QSO-Galaxy cross-correlations}

In order to check for the existence of QSO-Individual galaxy gravitational
lensing effects on the sample, we have to extract information on 
sub-degree angular distances, taking away the contribution from the
large scale galaxy distribution. We have used two independent methods
that agree in finding no correlation at these distances with a
$\approx 100$\% confidence level. We have excluded quasars with 
redshift $z<0.4$ to make sure the only possible correlations on these
short angular distances come from line-of-sight effects such as gravitational
lensing or dust absorption. A final list with 146 QSOs is thus considered.

\subsection{Nearest Neighbor Estimation}
Small scale correlations are harder to check in this work: The list
does not have many quasars which means poor statistics. Besides, there
is only one region over which the quasar catalog is defined. Hence  we
will be adding the same galaxies many times when computing correlations.
Gradients in the galaxy density might yield bogus QSO-galaxy
correlations. It is thus necessary to cross-check the actual sample
with random QSO catalogs which are roughly distributed the same way
on large scales as the original survey. Following previous work on the
subject (Thomas, Webster \& Drinkwater 1995)\markcite{to95}, 
we chose the closest QSO-galaxy distance
as the estimator of correlations on short scales. A histogram with
these distances for each quasar can be readily compared with the same
one for a random sample made by shifting 10 arcminutes the original survey
four times: North, South, East and West. A rank correlation test comparing
these two distributions finds no difference between them with a 
$\approx 100$\% confidence level (see Figure 4). 
We applied the same treatment to the bright
($B_j<19.6$) and faint ($B_j>19.6$) QSO subsamples. Gravitational lensing
predicts these two catalogs should have different short-scale correlations
because of the different slope in the number counts function. No correlation
was found for either case at a $\approx 100$\% confidence level.

To double check this result, we considered a list comprising all QSOs in the
galaxy region (289 deg$^2$) from the catalog of Veron-Cetty \& Veron
(1996). \markcite{vv96} A total of 508 QSOs were used and no excess 
or defect was found again at $\approx 100$\% confidence. Even though this
catalog is not complete, we do not anticipate a strong difference from a
complete sample since this estimator only has to do with short angular
distances and Veron's list compiles many different catalogs which 
should not imply a bias on such scales. It would be very
interesting to follow the same steps only for radio-loud quasars 
as previous studies found evidence for positive correlations 
explained by a double magnification bias effect. 
However, it is not possible to do this because  the list has only
30 radio--loud quasars inside this area, too few to infer a statistically
significant value. 

\subsection{Sub-degree Correlation Estimator}
Another way of checking for possible cross-correlations on small 
angular distances is defining a correlation estimator that is similar
to the correlation function used above. The actual sample, though, is
compared with a random sample of QSOs that preserve the large scale
structure of the original catalog. It is obtained by dividing the region
in $6\times 6$ boxes ($\sim 25^\prime \times 22^\prime$) and 
randomly throwing inside each box the same number of QSOs as the ones found
in the original sample. The correlation estimator is computed the same way
as the correlation function defined above (equations 1 and 2).
The result is shown in Figure 5 using again 200 simulated catalogs: 
The error bars are compatible with
no cross-correlation for either the whole catalog or partial subsamples,
in agreement with the outcome from the nearest neighbor histogram.

\section{DISCUSSION}

The effect of gravitational lensing on short angular distances on
this subsample of the optically-selected CFHT/MMT
survey can be seen in Figure 4.  There is no significant 
correlation on angular distances $\theta\lesssim 10^\prime$. This result 
was further checked with all the quasars listed in the 
Veron-Cetty \& Veron (1996)\markcite{vv96} catalog, 
for which no correlation was found either. The overdensity factor $q$ 
depends on the lens magnification $\mu$ and the slope of the QSO number 
counts function $\alpha$ as:
\begin{equation}
q \propto \mu^{\alpha -1}
\end{equation}
The slope of the number counts function determines the correlation:
positive for a steep slope ($\alpha > 1$) and negative for a shallow
slope ($\alpha < 1$).  
Figure 6 shows the number counts function versus
$B_j$ photographic magnitude and redshift for the complete CFHT sample
(top panels) and separated in two populations lying on the high and
low galaxy density areas respectively (bottom panels). The complete sample
has two slopes: The bright section has $\alpha = 1.7$ (positive correlation
expected) whereas the faint section has $\alpha = 0.7$ (slightly negative
correlation). Separating the samples at the point where the slope
changes (around $B_j \sim 19.5$) allows us to estimate whether the
lensing effect is noticeable. No correlation was found, with a
$\approx 100$\% confidence level using a Spearman rank correlation test. 

The new issue that arises in this work has to do with possible biasing
in optically selected samples. We were lucky to use the best QSO survey
that minimizes selection effects (line emission). Yet, we found a
very significant QSO--galaxy anticorrelation on large scales that --- on a
first estimate --- could be associated with dust tracing the distribution
of foreground galaxies. Figures 2 and 3 show this negative correlation 
that can be
even checked by eye in Figure 1. However, when we take into account
different subsamples, we find the low redshift population is the one
responsible for this negative correlation, as can be shown in the bottom 
panels of Figure 2.
Selecting a subsample of low-$z$ quasars and red galaxies
(which trace most of the galaxy density gradient), we find a significant
anticorrelation with a $>99.6$\% confidence level. Taking a look at the
density--density diagrams, a defect of 50\% in the low-$z$-QSO number 
counts appears in high galaxy density regions. 
An estimation for the complete sample 
obtained either from the linear fits from Figure 3 or by integrating 
the correlation function in Figure 2 yields a quasar defect around 10\%.
That is, if the quasar list were truly complete, we would expect around 15
extra QSOs, most of them with redshifts $z<1.6$.

This effect can only be explained by a
redshift--dependent bias on the selection process. This has been already
considered by Hartwick \& Schade (1990) \markcite{hs90} but so far
it had not been possible to quantify. High redshift quasars are mainly
detected through their strong Lyman-$\alpha$ emission peak, whereas
low redshift objects must be detected using less prominent emission lines
that may introduce this bias. If we assume a spectral sensitivity
range of $\lambda\lambda$ 3500--5400 $\AA$ for the CFHT survey with III-aJ
plates (Crampton, Schade \& Cowley 1985) \markcite{cr85}, we get a 
redshift range $1.9 < z < 3.4$ using the Lyman-$\alpha$
peak; $1.3 < z < 2.5$ using \ion{C}{4}; and $0.25 < z < 0.9$ using
\ion{Mg}{2}. If we take into account the stronger flux emitted in the
Lyman-$\alpha$ peak, we may expect to find more QSOs at high redshift
using this selection process. Hence, as opposed to earlier claims,
the blue grens surveys could be significantly susceptible to the
``Ly-$\alpha$ window bias'' altering the completeness level 
of the catalogs obtained this way. Besides, we have found that the selection 
method is less efficient around high galaxy density areas, where the
crowded grens plates might make searching for faint emission lines 
quite a challenge. This result should be explored further using QSO 
surveys on regions with a strong galaxy density gradient (that is 
essential because the anticorrelation obtained in this work
is only significant taking the red galaxies that trace the large scale
structure). The important implications that this correlation has on
the completeness level of surveys as well as on any study dealing
with quasar--galaxy associations make the effort worthwhile.

\acknowledgments
We are happy to thank S.C.Odewahn \& G.Aldering for sending us a 
machine-readable file of their galaxy catalog. I.F. and N.B. acknowledge 
a Ph.D. scholarship from the `Gobierno de Cantabria', and
the Spanish MEC, respectively. I.F.,N.B. and E,M.-G. acknowledge financial 
support from the Spanish DGES under contract PB95-0041.

\newpage

\figcaption[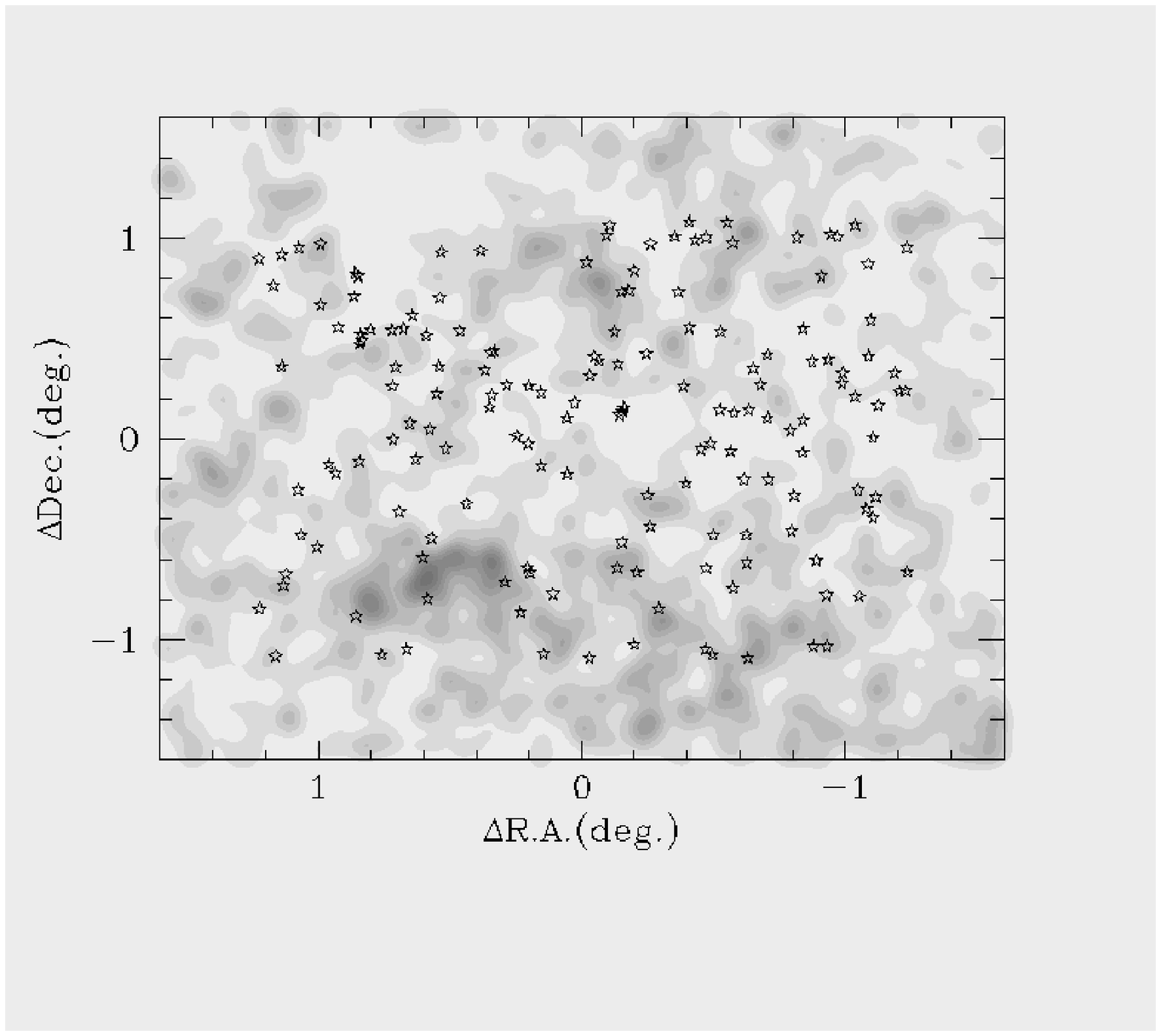]{Region where the QSO catalog used is complete
($\approx 5.5$deg$^2$). 
North and East are up and left respectively. The center corresponds to 
R.A.=$13^h37^m13.25^s$ ; Dec.= $+27^o20^\prime 00^{\prime\prime}$
(B1950.0). The 160 quasars are shown as stars and the galaxies are
smoothed by convolution with a gaussian (FWHM$\approx 1^\prime$,
roughly the average galaxy--galaxy separation)
into a density function shown with a grey scale. Notice the
paucity of quasars around the high density filament in the bottom part of
the figure. \label{fig1}}

\figcaption[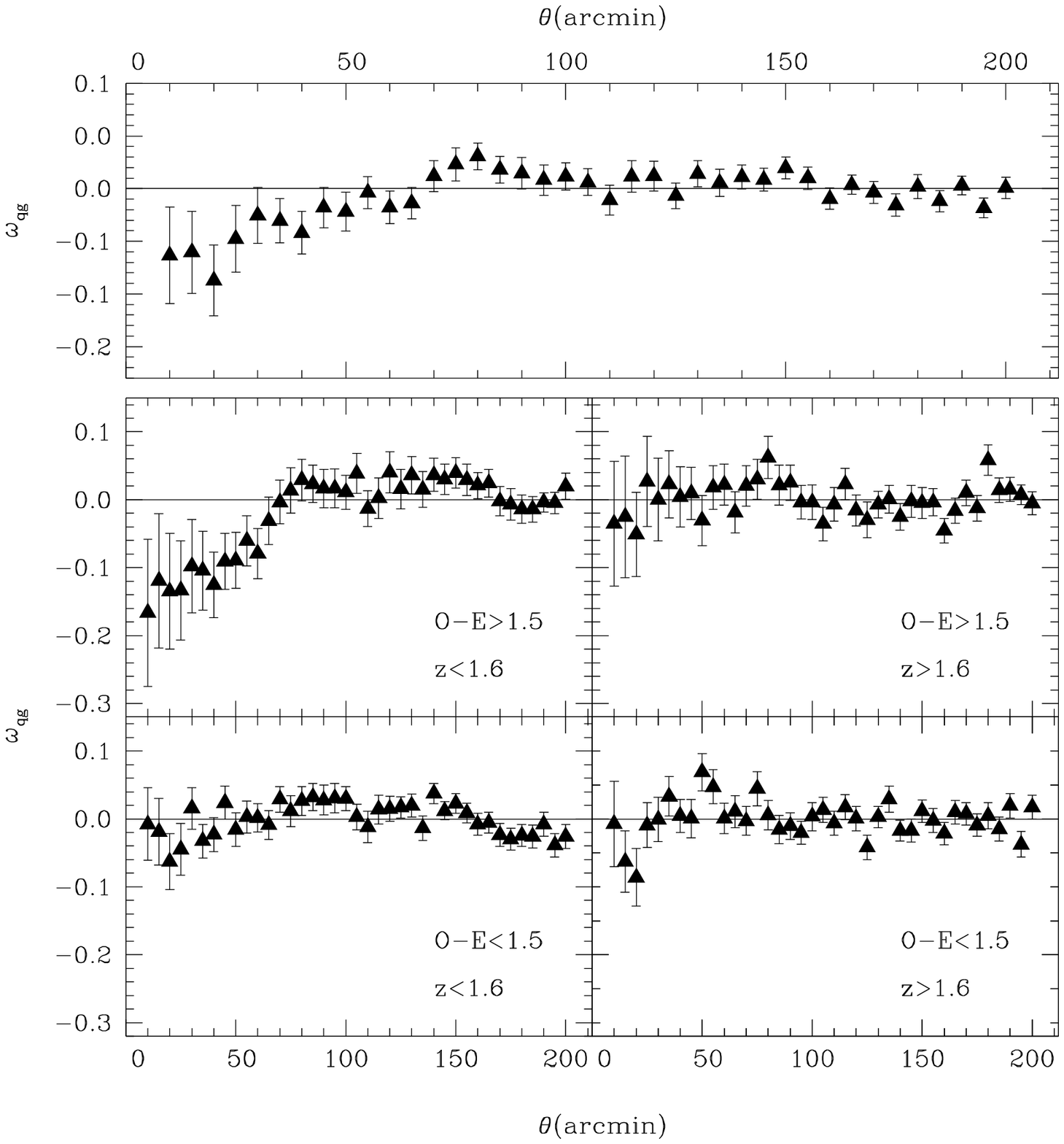]{Cross--Correlation Function. 1-$\sigma$
error bars are obtained by comparing the real QSO sample with 200 
random catalogs with the same number of quasars. The subsample comprising
red galaxies and low redshift quasars gives a non--zero cross--correlation
function suggesting a bias in grism--selected surveys
against detection of low redshift quasars around high galaxy density. 
A Spearman Rank correlation test gives a $\approx$ 100 \% confidence
level. This
bias can be due to the need to detect fainter emission lines in contrast
with high-$z$ QSOs for which the strong Lyman-$\alpha$ peak falls within
the spectral range of the detector.\label{fig2}}

\figcaption[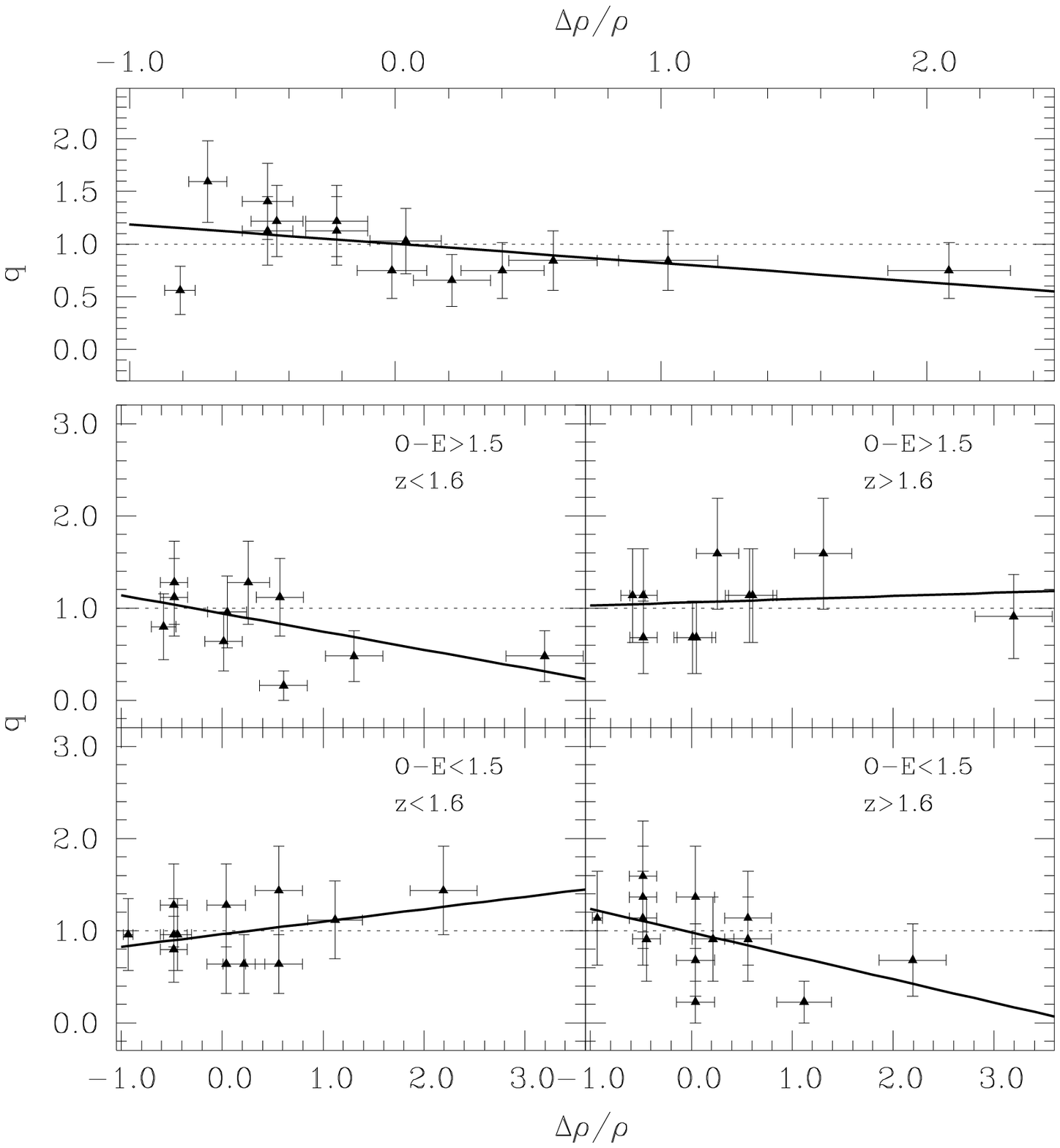]{Density--density 
plots for different samples: The galaxy density
gradient ($\Delta\rho /\langle \rho\rangle$) is plotted versus QSO number
(q=$N_q/\langle N_q\rangle$). Top panel: Complete sample of QSOs and galaxies. 
The bottom panels show the same plot for a few subsamples: red($O-E>1.5$) 
and blue galaxies and high and low redshift QSOs. The line gives a 
least squares fit to each diagram. Notice the agreement of this method with
the cross--correlation function shown in figure 2: the
subsample with red galaxies (which trace the gradient in the galaxy density)
and low--redshift quasars is the one with a significant negative 
cross--correlation. The remaining panels give a result compatible within 
error bars with no correlation whatsoever.\label{fig3}}

\figcaption[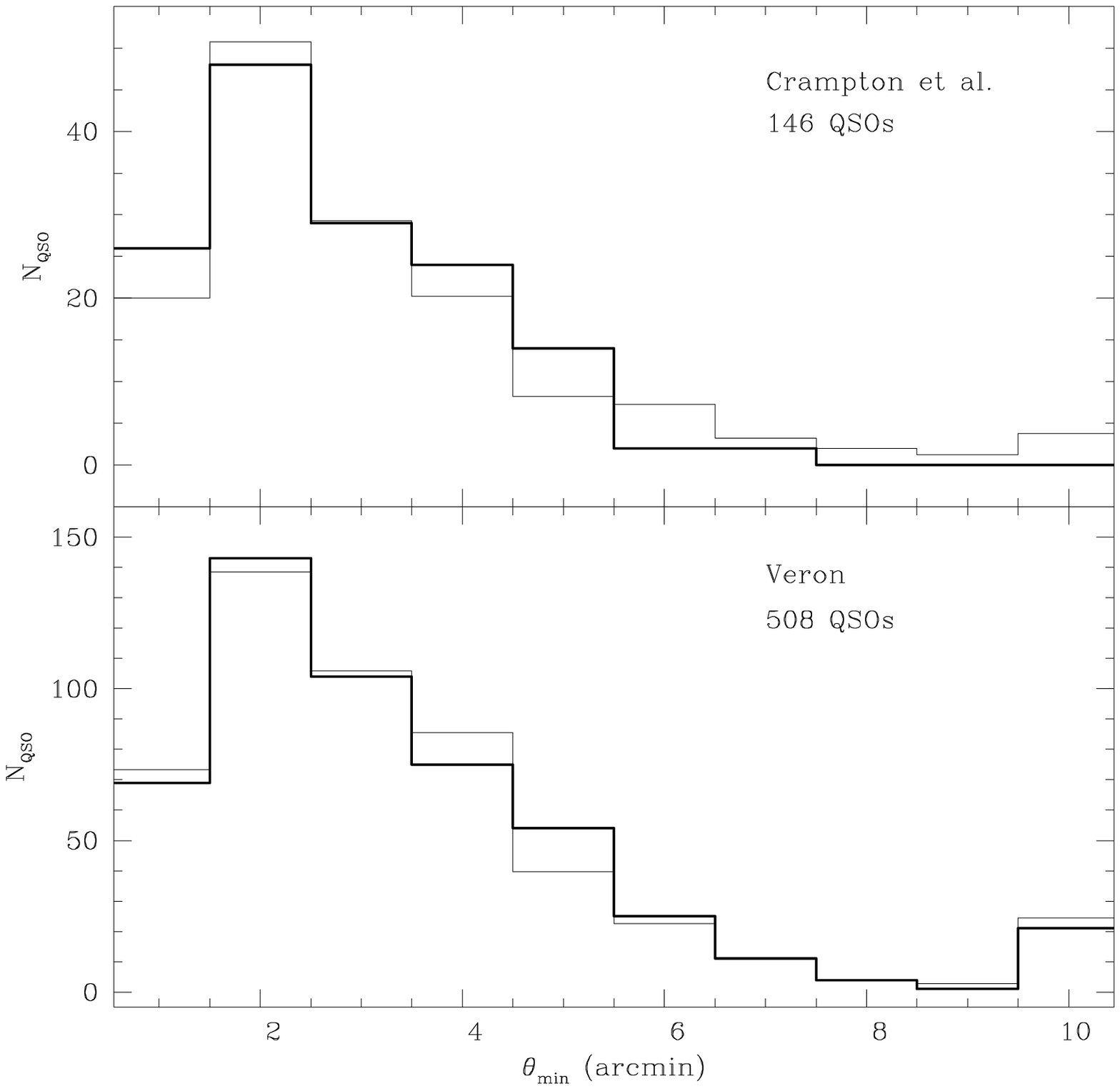]{Nearest neighbor estimator. The top panel shows the 
result for the QSO catalog used in this work. The bottom panel serves as 
a check using all quasars found in the list from Veron-Cetty \& Veron (1996) 
lying inside 
the 289 deg$^2$ galaxy region. Thick lines represent the actual sample 
whereas thin lines show the random catalogs. The result is compatible
with no cross--correlation on sub--degree angular distances (confidence
level $\approx 100$\% using a rank order test).
\label{fig4}}

\figcaption[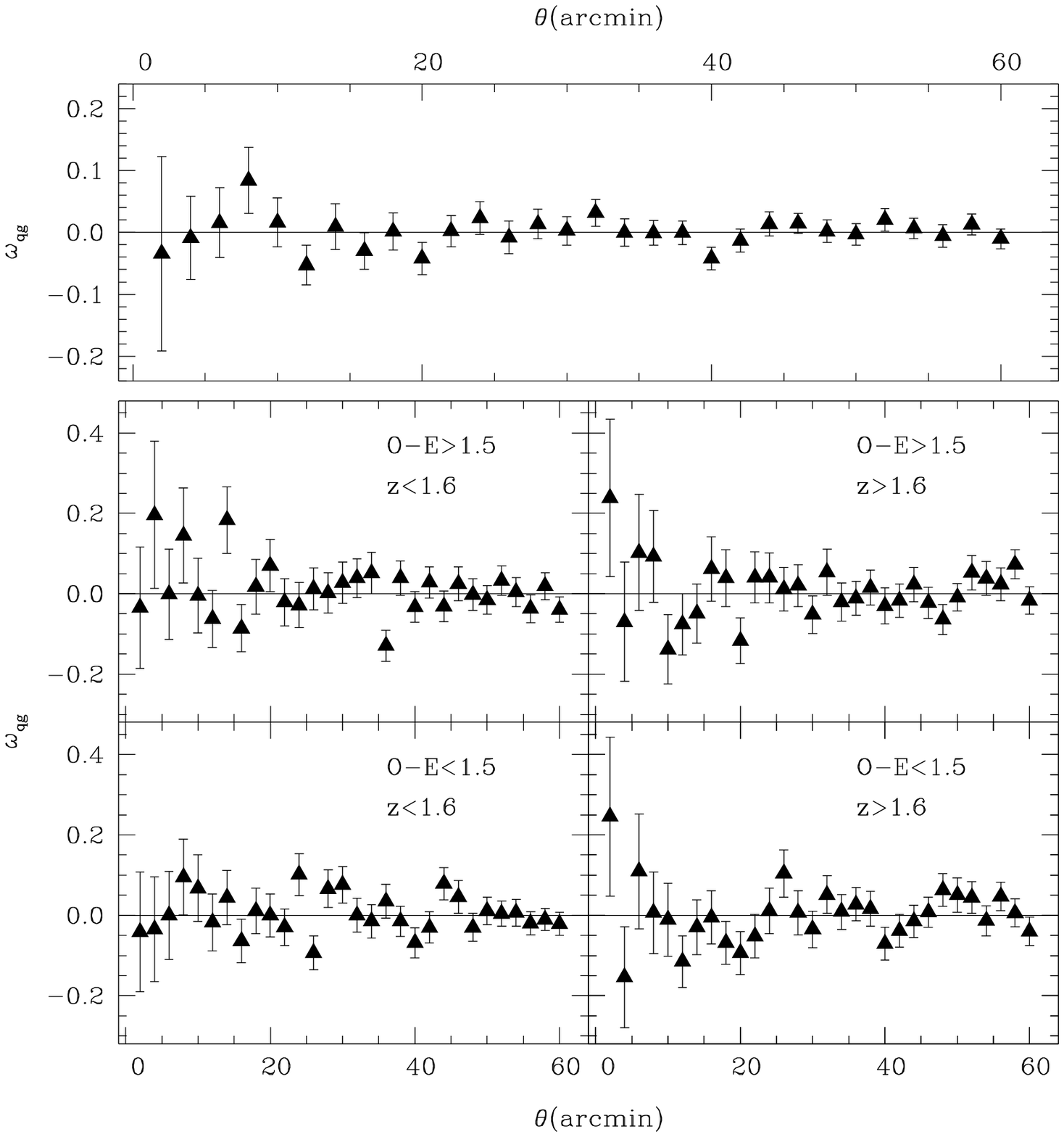]{Correlation estimator for small angular distances
using 200 random catalogs to match against the real sample. 
1-$\sigma$ error bars are shown. the top panel
shows the result for the complete catalogs. Bottom panels show the function
for different subsamples. This method is totally independent from a
nearest neighbor estimator (fig.4), yet it agrees in finding no
cross--correlation for any subsample.
\label{fig5}}

\figcaption[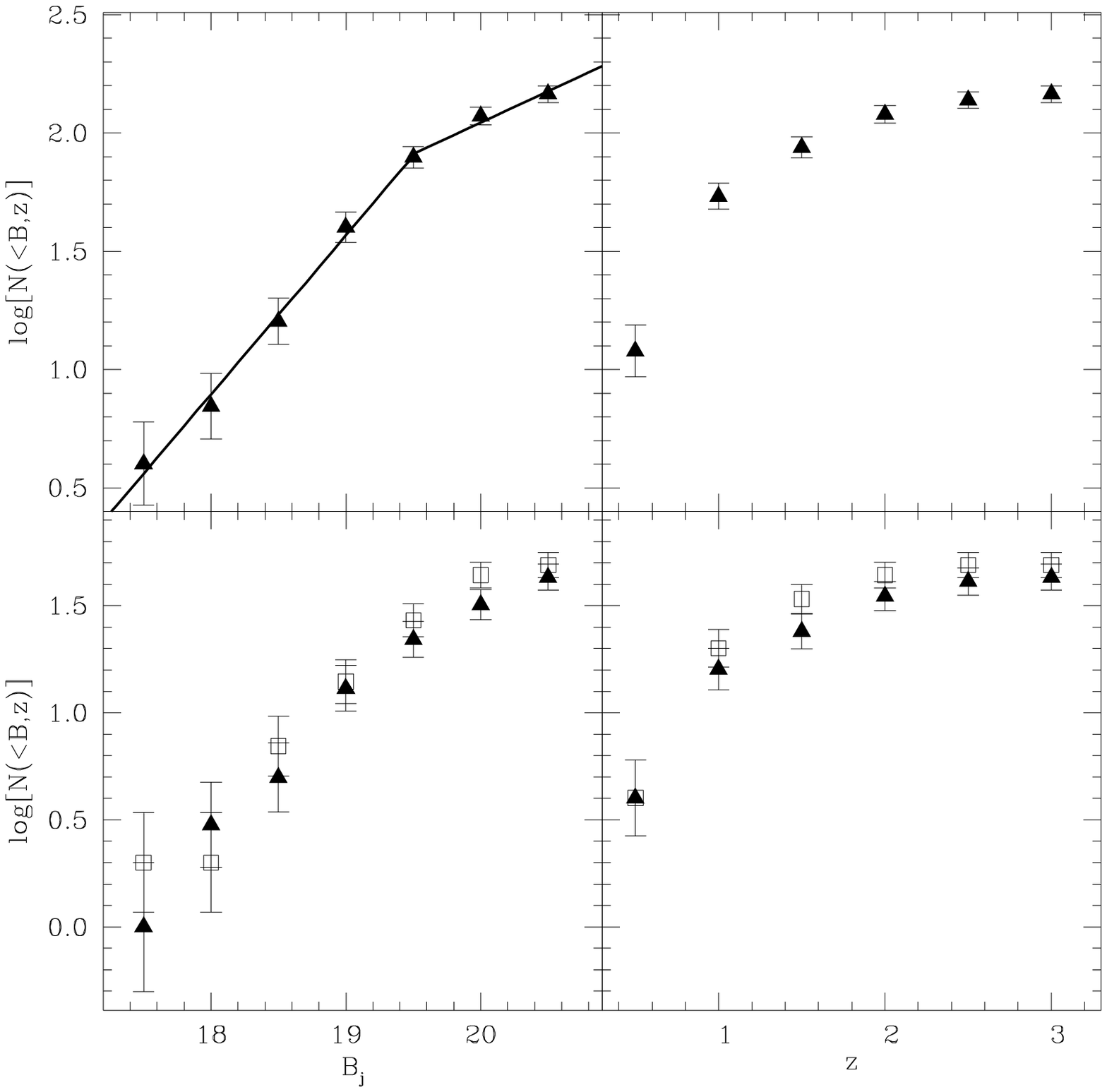]{QSO number counts versus magnitude and redshift.
The top panels show the functions for the complete sample; the bottom 
panels take the high (triangles) and low (squares) galaxy density regions
separately. Poisson error bars are shown. The slopes of the linear fit
in the number counts versus magnitude diagram are 1.7 and 0.7 for
$B_j\le 19.5$ and $B_j>19.5$ respectively.\label{fig6}}

\end{document}